# Sistem Informasi Geografis Potensi Wilayah Kabupaten Banyuasin Berbasis Web

Rastuti[1], Leon Andretti Abdillah[2], Eka Puji Agustini[3]
[1,2,3] Program Studi Sistem Informasi, Fakultas Ilmu Komputer, Universitas Bina Darma
Palembang, Indonesia
[1]rastuti_15@yahoo.co.id, [2]leon.abdillah@yahoo.com

**Abstract.** Development of information technology in the field of spatial data processing has helped many digital mapping. In the present study, the authors will empower geographic information systems (GIS) for geographic data processing potential Banyuasin district. The potential of the region are agriculture, farming, and industry. The method used to develop the GIS is the waterfall approach models. After conducting a series of activities starting from the analysis, requirements, system design, coding, testing, then obtained an information systems that can provide information about the geographical spread of the potential of web-based Banyuasin district with the help of ArcGIS. This system was built by using the waterfall model. The result is a GIS that provide location information potential of the region.

## 1 Pendahuluan

Teknologi informasi (TI) merupakan bagian terpenting dari kebutuhan dalam memberikan suatu informasi yang dibutuhkan oleh pengguna baik untuk menyimpan, mengelola dan menganalisis serta memanggil data. Agar data yang dibutuhkan tersebut menjadi lebih efektif dan efisien, salah satunya pemanfaatan dalam sistem informasi geografis (SIG). Dengan memanfaatkan SIG akan memberikan kemudahan kepada para pengguna atau para pengambil keputusan untuk menentukan kebijaksanaan yang akan diambil [1]. SIG adalah suatu sistem berbasis komputer untuk menangkap, menyimpan, mengecek, mengintegrasikan, memanipulasi, dan men-*display* data dengan peta digital [2]. SIG sudah digunakan secara luas untuk mengakses informasi tentang suatu lokasi [3]. Keputusan yang diambil khususnya yang berkaitan dengan aspek keruangan/spasial [4]. Pada penelitian ini, penulis akan memanfaatkan teknologi SIG untuk data lokasi potensi wilayah.

Potensi wilayah Kabupaten Banyuasin akan berkembang baik, bila pertumbuhan potensi seperti pertanian, perindustrian dan perkebunan dikelola dengan baik, sehingga akan memberikan kontribusi pendapatan ekonomi yang semakin meningkat. Untuk dapat mengetahui informasi perbandingan wilayah administrasi Kabupaten Banyuasin diperlukan suatu SIG dalam penyebaran potensi yang dapat direalisasikan melalui teknologi sistem informasi geografis berbasis *web*. Tujuan dari penelitian ini adalah membuat SIG kepada pengguna yang terdiri dari kecamatan, perusahaan dan instansi-intansi terkait yang memerlukan informasi tentang letak potensi wilayah di bidang pertanian, perkebunan dan perindustrian yang ada di Kabupaten Banyuasin.

Dalam penelitian ini penulis akan memetakan potensi wilayah pada Kabupaten Banyuasin Sumatera Selatan yang bergerak pada bidang pertanian, perindustrian dan perkebunan. Kabupaten Banyuasin terbagi atas 19 (sembilan belas) kecamatan [5]. Peta administrasi Kabupaten Banyuasin [6] dapat dilihat pada gambar 1. Setiap kecamatan tersebut memiliki potensi alam yang melimpah yang sebaiknya disajikan dalam bentuk sistem informasi geografis berbasis *web*.

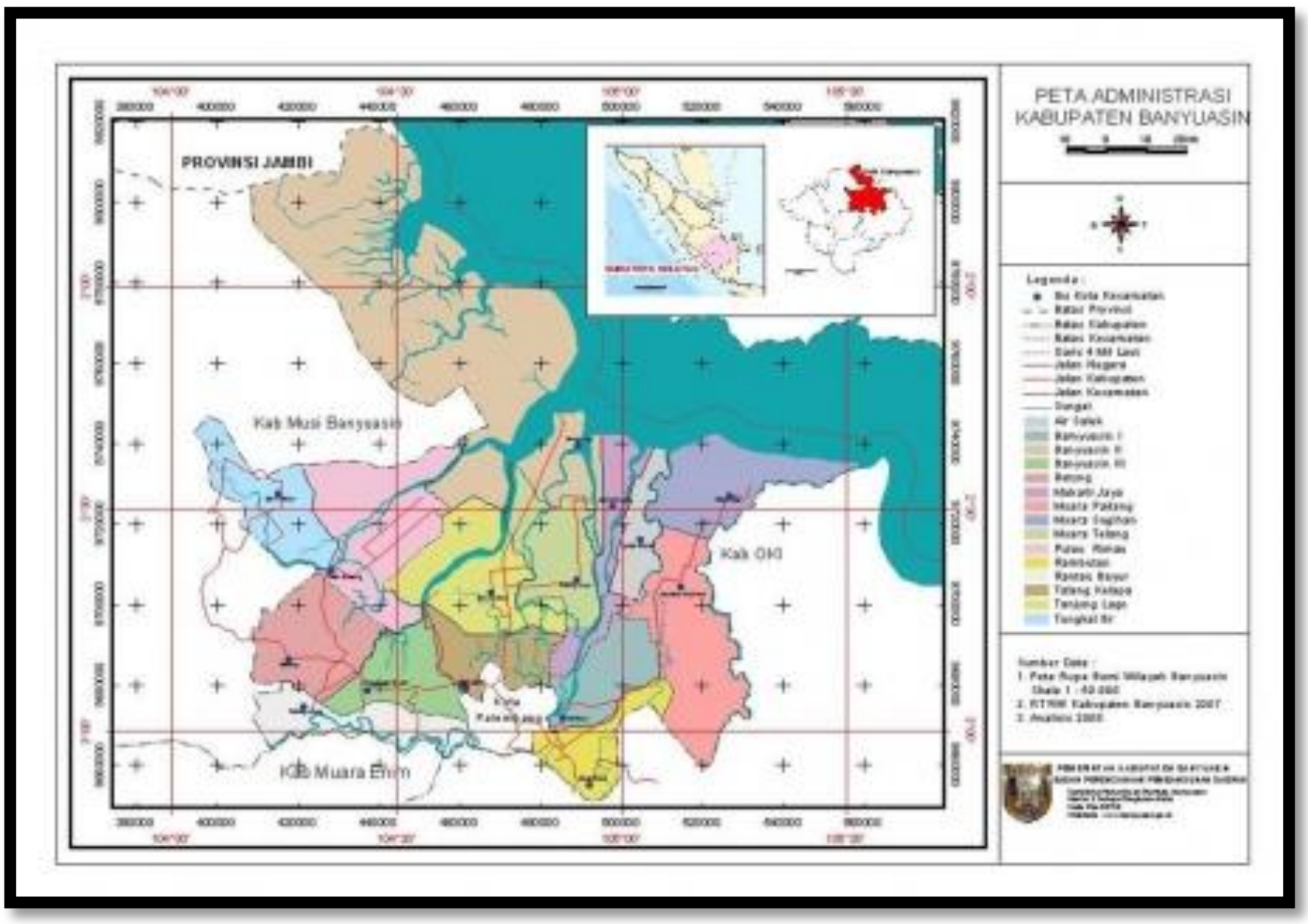

**Gambar 1.** Peta Administrasi Kabupaten Banyuasin

Sejumlah penelitian telah penulis kaji untuk pengembangan SIG ini, antara lain: 1) Sistem informasi geografis lokasi perumahan berbasis android [3], dan 2) Aplikasi Sistem Informasi Geografis dalam Pemetaan Batas Administrasi, Tanah, Geologi, Penggunaan Lahan, Lereng, Daerah Istimewa Yogyakarta dan DAS di Jawa Tengah Menggunakan Software ArcView GIS [7].

## 2 Metode Penelitian

Data yang digunakan pada penelitian ini didapat dengan: 1) proses wawancara yang dilakukan dengan pihak BAPPEDA Banyuasin untuk mendapatkan gambaran kebutuhan pengguna terhadap sistem yang akan dibangun, dan 2) studi literatur berupa penelusuran jurnal ilmiah dna dokumentasi yang ada.

Metode yang digunakan dalam pembangunan perangkat lunak ini adalah model air terjun (*waterfall model*) atau sering disebut dengan "siklus kehidupan klasik". Tahapan utama dari *waterfall model* langsung mencerminkan aktifitas pengembangan dasar. Terdapat 5 (lima) tahapan pada *waterfall model* [8], yaitu: 1) *requirement analysis and definition*, 2) *system and software design*, 3) *implementation and unit testing*, 4) *integration and system testing*, dan 5) *operation and maintenance*.

## 3 Hasil dan Pembahasan

Setelah melakukan kegiatan analisis, dan rekayasa sistem yang telah dibahas sebelumnya, maka hasil yang diperoleh dalah sebuah SIG potensi wilayah kabupaten Banyuasin berbasis *web*. Peta hasil pembuatan sistem informasi geografis kepada pengguna yang terdiri dari wilayah kecamatan, perusahaan dan instansi-intansi terkait yang memerlukan informasi tentang letak potensi wilayah di bidang pertanian, perkebunan dan perindustrian yang ada di Kabupaten Banyuasin.

Hasil dari sistem ini memudahkan pengguna yaitu kecamatan, perusahaan dan instansi-instansi untuk mengetahui informasi potensi dibidang pertanian, perkebunan dan perindustrian pada setiap wilayah Kabupaten Banyuasin. Selain itu juga memberikan kemudahan petugas dalam mengatur pengolahan data potensi wilayah.

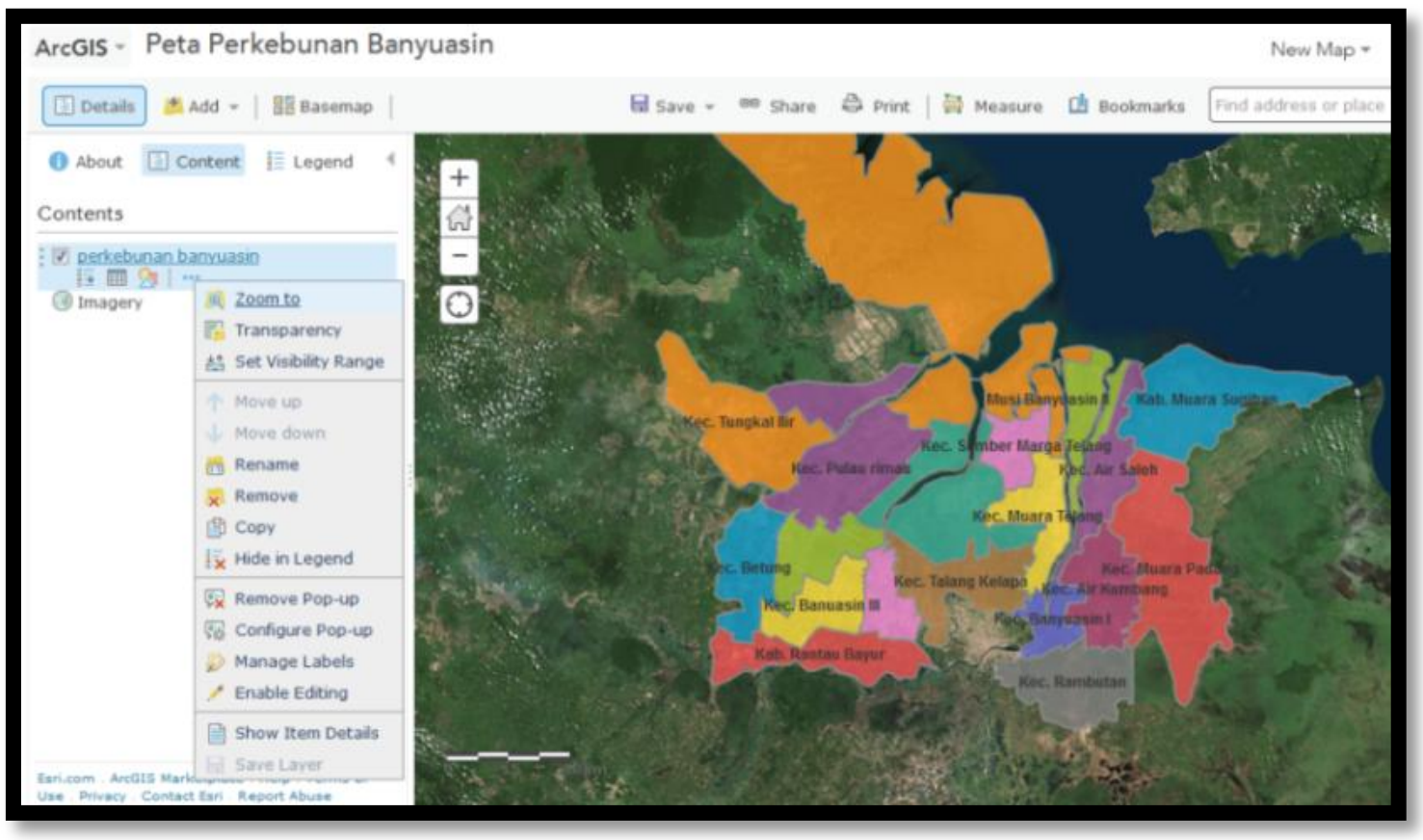

**Gambar 2.** Halaman *Upload ArcGIS Online*

### 3.1 Halaman WebGIS Pertanian

Setelah proses peletakan ArcGIS secara *online* selanjutnya yaitu proses *embeded* data peta tersebut ke WebGIS yang yang dibuat. Halaman berikut menampilkan informasi tentang *website* dengan peta ArcGIS yang telah tertanam (*embedded*). Halaman ini menggambarkan informasi mengenai informasi pertanian jika kita mengklik wilayah kecamatan maka akan menampilkan informasi detail mengenai wilayah tersebut.

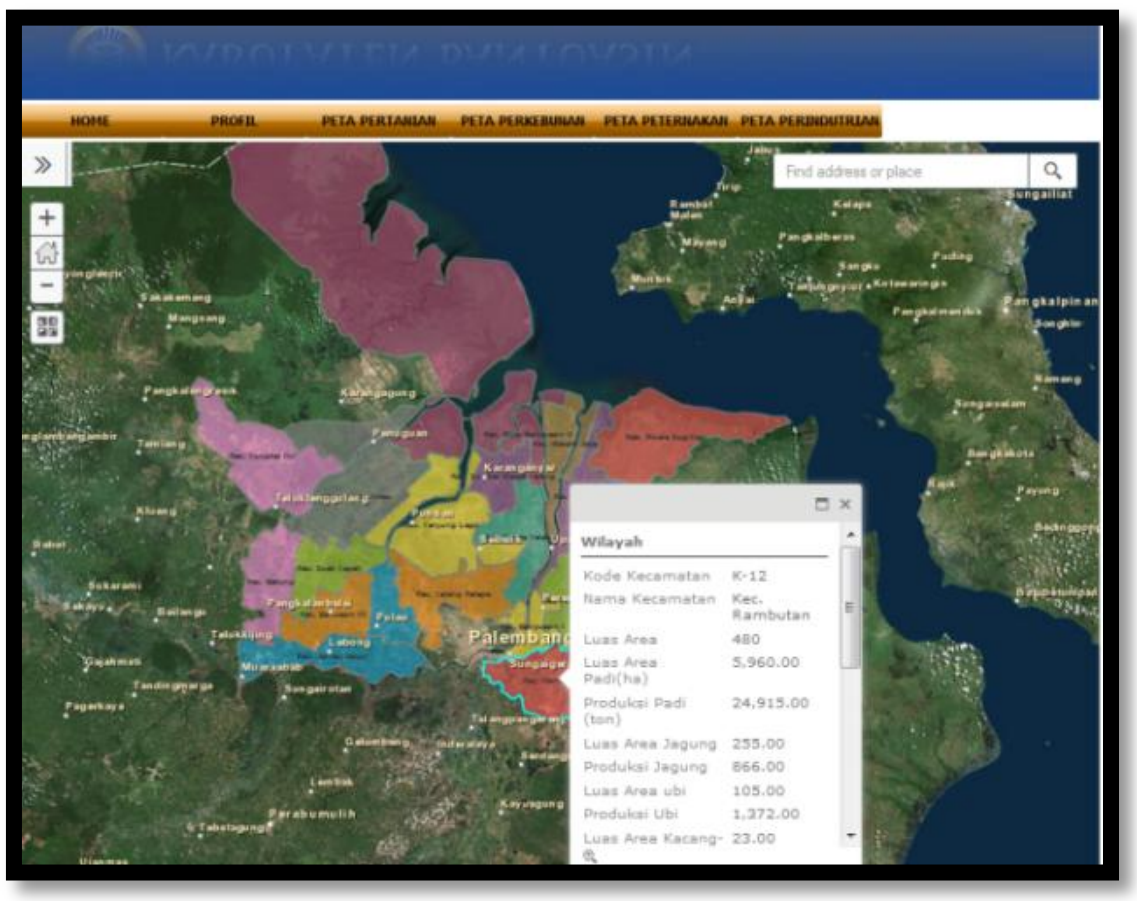

**Gambar 3.** WebGIS Peta Pertanian

### 3.2 Halaman WebGIS Perkebunan

Halaman ini menggambarkan informasi mengenai informasi perkebuan jika kita mengklik wilayah kecamatan maka akan menampilkan informasi detail mengenai perkebunan yang ada di wilayah tersebut (gambar 4).

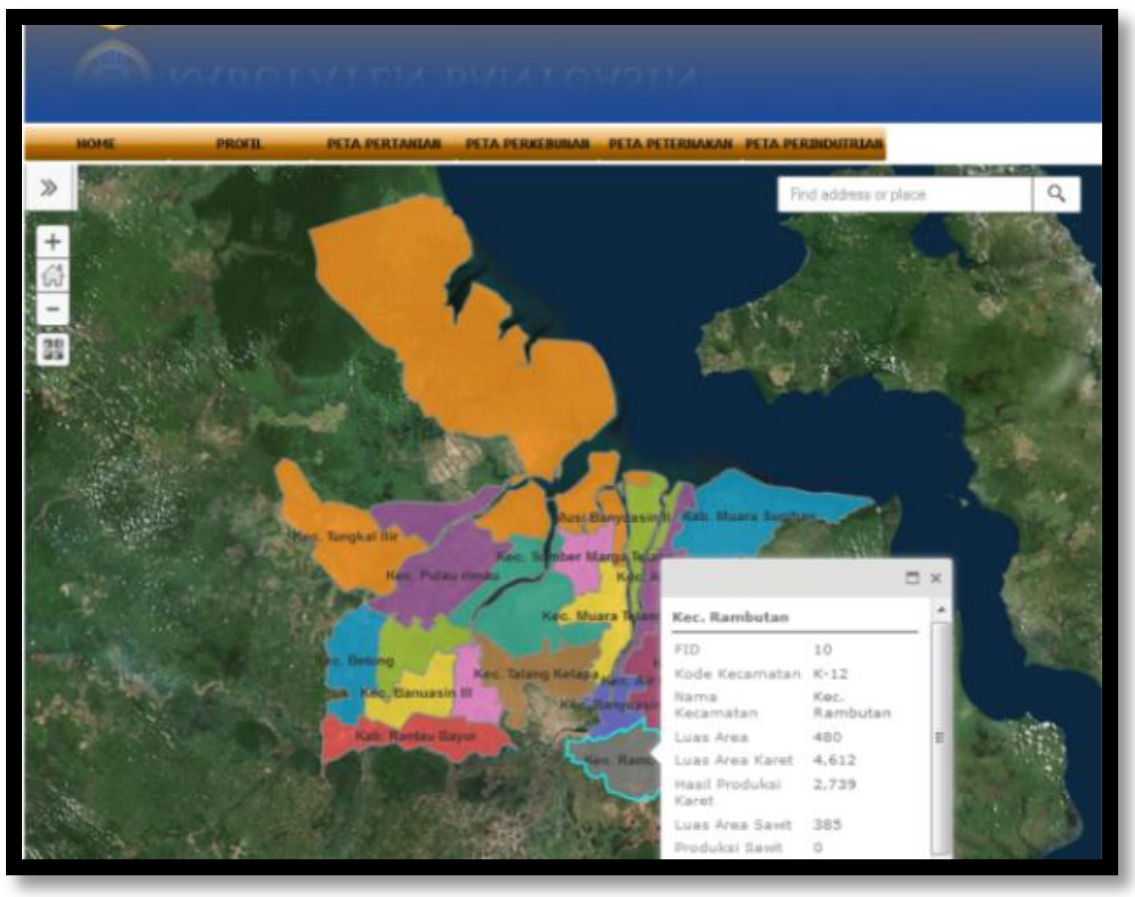

**Gambar 4.** WebGIS Peta Perkebunan

### 3.3 Halaman WebGIS Perindustrian

Halaman ini menggambarkan informasi mengenai informasi industri jika kita mengklik wilayah kecamatan maka akan menampilkan informasi detail mengenai indutri yang ada di wilayah tersebut (gambar 5).

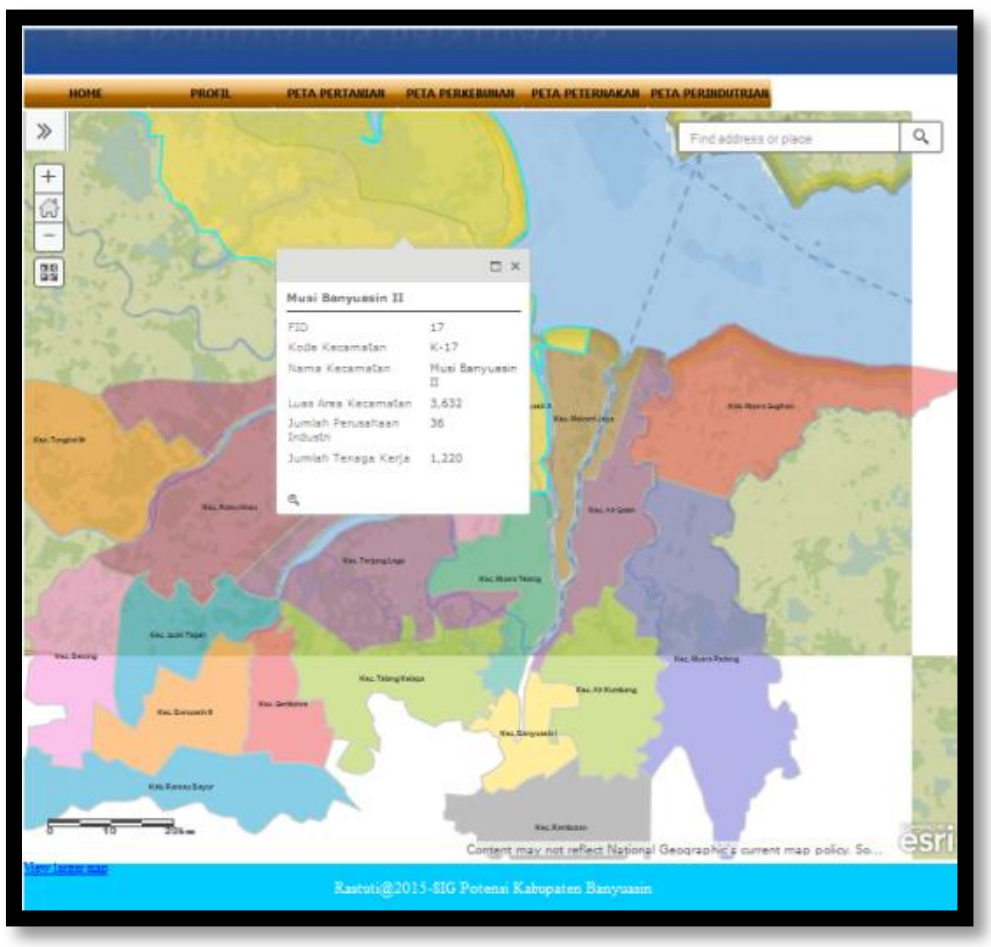

**Gambar 5.** WebGIS Peta Perindustrian

## 4 Kesimpulan

Berdasarkan analisis, perancangan dan pengembangan sistem, serta ujicoba, maka dapat disimpulkan:

1. SIG ini memudahkan pengguna untuk mengetahui informasi potensi dibidang pertanian, perkebunan dan perindustrian di Kabupaten Banyuasin.
2. Memberikan kemudahan kepada petugas dalam mengatur pengolahan data potensi wilayah di Kabupaten Banyuasin.
3. Manfaat dalam pengolahan data potensinya agar lebih memudahkan pengembangan wilayah Kabupaten Banyuasin.
4. Untuk pengembangan selanjutnya, SIG ini dapat dikembangkan dengan dilengkapi oleh fasilitas penyimpanan hasil pencarian ke format PDF [9] sebagai standar dokumen global.

## Daftar Pustaka